\documentclass[%
 reprint,
 amsmath,amssymb,
 aps,
 floatfix,
]{revtex4-2}

\usepackage{graphicx}
\usepackage{dcolumn}
\usepackage{bm}

\usepackage{braket}
\usepackage{tikz}
\usepackage{float}
\usepackage{todonotes}

\usepackage{algorithm}
\usepackage{algorithmicx}
\usepackage{algpseudocode}

\newcommand{\myeq}[1]{\begin{eqnarray}\begin{aligned}#1\end{aligned}\end{eqnarray}}

\begin{document}

\preprint{APS/123-QED}

\title{Efficient quantum readout-error mitigation for\\
sparse measurement outcomes of near-term quantum devices}

\author{Bo Yang$^1$}
\author{Rudy Raymond$^{2,3}$}
\author{Shumpei Uno$^{3,4}$}
\affiliation{%
 $^1$Graduate School of Information Science and Technology, The University of Tokyo, Bunkyo-ku, Tokyo 113-8656, Japan\\
 $^2$IBM Quantum, IBM Research-Tokyo, 19-21 Nihonbashi Hakozaki-cho, Chuo-ku, Tokyo, 103-8510, Japan\\
 $^3$Quantum Computing Center, Keio University, Hiyoshi 3-14-1, Kohoku-ku, Yokohama 223-8522, Japan\\
 $^4$Mizuho Research \& Technologies, Ltd, 2-3 Kanda-Nishikicho, Chiyoda-ku, Tokyo, 101-8443, Japan
}%


\date{\today}

\begin{abstract}

The readout error on near-term quantum devices is one of the dominant noise factors, which can be mitigated by classical postprocessing called quantum readout error mitigation (QREM).
The standard QREM applies the inverse of noise calibration matrix to the outcome probability distribution using exponential computational resources to the number of measured qubits.
This becomes infeasible for the current quantum devices with tens of qubits or more.
Here we propose two efficient QREM methods finishing in $O(ns^2)$ time for probability distributions of $n$ qubits and $s$ shots, which mainly aim at mitigating sparse probability distributions such that only a few states are dominant.
We compare the proposed methods with several recent QREM methods in the following three cases: expectation values of the GHZ state, its fidelities, and the estimation error of maximum likelihood amplitude estimation (MLAE) algorithm with a modified Grover iterator.
The two cases of the GHZ state are on real IBM quantum devices, while the third is with numerical simulation.
Using the proposed method, the mitigation of the 65-qubit GHZ state takes only a few seconds, and we witness the fidelity of the 29-qubit GHZ state exceeding 0.5.
The proposed methods also succeed in reducing the estimation error in the MLAE algorithm, outperforming the results by other QREM methods in general.
\end{abstract}

\maketitle



\section{\label{sec:introductions}Introduction\protect\lowercase{}}


\par The quantum information on quantum devices is vulnerable to various noises due to the incompleteness of the quantum state or unwanted interactions with outside world.
Since the significance of quantum computing has been recognized, many quantum error correction methods were proposed for the protection of quantum information \cite{PhysRevA.52.R2493, PhysRevLett.77.793, Kitaev_2003, Chamberland_2020}.
However, the current and near-term quantum devices are so small and noisy that those prominent methods are not yet applicable.
Nevertheless, one can mitigate the errors occurring in the quantum process to obtain meaningful results on the current and near-future quantum devices.
Error mitigation aims to directly retrieve noiseless results by adding mitigation gates to quantum circuits or performing classical postprocessing after measurements, such as zero-noise extrapolation \cite{Temme_2017, Kandala2019, Giurgica_Tiron_2020}, probabilistic error cancellation \cite{Temme_2017, takagi2020optimal}, dynamic decoupling \cite{Wei2019, Pokharel_2018}, readout error mitigation for expectation value \cite{ctmp2020bravyi, berg2021modelfree, chen2021, hicks2021rebalancing}, and many new methods for various noise models \cite{McClean2020, yoshioka2021generalized, endo2021hybrid, endo2019hamiltonian, czarnik2021learning, strikis2021learning, sun2021mitigating, otten2018accounting}.
These error mitigation techniques are often combined with near-term quantum algorithms \cite{vqe2014, qaoa2014quantum}.


\par Here we focus on the mitigation of quantum readout error, one of the significant noise factors on current near-term devices.
When the state preparation noise is minimal, the readout error can be characterized by a stochastic matrix called the calibration matrix, whose elements represent the transition probability from the expected measured states to the actual measurement outputs \cite{Lundeen2009, Maciejewski2020mitigationofreadout}.
The quantum readout error mitigation (QREM) performs classical postprocessing by applying the inverse of the calibration matrix to the measured probability distribution (i.e., the frequency distribution of measured bitstrings).
Since the number of possible quantum states is $2^n$ for measurement of $n$-qubit system, rigorous inversion of this calibration matrix requires exponential time and memory on classical computers, which is infeasible for the measurement results from the current and near-future quantum devices with tens and hundreds of qubits.

\par Towards this issue, several scalable approaches have been already proposed \cite{mooney2021generation,nation2021}.
These methods assume the readout noises follow the tensor product noise model, where the readout noise on each qubit or qubit block is considered local.
Under this assumption, Mooney \textit{et al.} \cite{mooney2021generation} sequentially apply the inverse of each small calibration matrix and cut off the vector elements smaller than the arbitrary threshold $t$.
While their method is practically very fast, a particular gap from the exact inversion result would be involved.
Also, its time complexity and the space complexity are not theoretically bounded.
Using this method, they witnessed the genuine multipartite entanglement (GME) of large GHZ states on the IBM Quantum device up to size 27 \cite{mooney2021generation}.

\par The other efficient approach by Nation \textit{et al.} \cite{nation2021} restricts the size of the calibration matrix to the subspace of measured probability distribution and applies the inverse of the reduced calibration matrix by matrix-free iterative methods.
This assumption is justified when the measured probability distribution contains a few principal bitstrings with high probability.
This method, named ``mthree" (matrix-free measurement mitigation), mitigates the measurement result of a 42-qubit GHZ state from the IBM Quantum device in a few seconds on a quad-core Intel i3-10100 system with 32 GB of memory with NumPy and SciPy compiled using OpenBlas.

\par Our proposed methods, which were conceived independently and whose preliminary results presented in~\cite{sigqs03,aqis2021}, are similar to the idea in mthree \cite{nation2021}.
We also assume the tensor product noise model and the reduced space of the calibration matrix.
The main difference lies in the order of matrix reduction and inversion.
The proposed methods directly compute each element in a reduced inverse calibration matrix while the reduction of calibration matrix comes first in \cite{nation2021}.
This is the most tedious step in the proposed method, taking $O(ns^2)$ time for the measurement result with $n$ qubits and $s$ shots.

\par In addition, the mitigated frequency vector must satisfy the property of probability distributions.
That is, the elements of the vector should be nonnegative and the element sum exactly becomes one.
Since the reduced inverse matrix would not preserve these conditions, we first make the sum of vector elements to one and next remove the negative elements.

\par The proposed method seems to be suitable for mitigating the probability distribution with a few dominant state probabilities.
To check the practical performance of the proposed methods, we conduct the following three demonstrations.

\par First, the expectation value of GHZ states is examined on the 65-qubit IBM Quantum Brooklyn superconducting quantum device provided by IBM Quantum Experience \cite{ibmq}.
The IBM Quantum Brooklyn has the Hummingbird r2 quantum processor and the heavy-hexagonal qubits structure.
Its quantum volume (QV) is 32.
With the C++/Cython implementation, the proposed methods mitigate the 65-qubit GHZ states on it within 5 s, which is practically fast enough.
The mitigation errors can be exactly computed in the proposed method, while the iterative methods by \cite{nation2021} would output only approximated values.

\par Second, the fidelity of GHZ states on IBM Quantum Brooklyn is computed.
The GME on the large quantum states (e.g., GHZ states and star graph states) has been widely investigated \cite{ghz_fidelity2017, ghz_fidelity2019, ghz_fidelity20, ghz_fidelity20mooney, Wei2019, mooney2021generation, yang2021testing}.
With the proposed efficient QREM, we witnessed the GME in the form of 29-qubit GHZ state on IBM Quantum Brooklyn with fidelity more than 0.5.

\par Furthermore, the estimation error of the maximum likelihood amplitude estimation (MLAE) algorithm \cite{Suzuki_2020} with a modified Grover iterator \cite{uno2020modified} is investigated by numerical simulation on Qiskit \cite{Qiskit}.
The amplitude estimation problem has essential applications in finance and machine learning using quantum devices \cite{giurgicatiron2020low, bouland2020prospects}.
In the noiseless environment, the estimation error of this modified Grover algorithm scales in the order of $O(1/N)$ for $N$ Grover oracle iterations, while even a small readout noise would spoil this convergence rate.
The proposed QREM methods succeed in recovering the original error convergence rate under readout noises on the noisy simulator.
While the modified Grover algorithm \cite{uno2020modified} is tolerant to depolarization errors, our numerical simulation suggests that it is also applicable under readout noises.

\section{\label{sec:propsed_method}Proposed Methods\protect\lowercase{}}
\subsection{Tensor Product Noise Model\protect\lowercase{}}
\label{sec:tensor_product_noise_model}
\par Under the complete noise model, the calibration matrix $A$ is obtained by examining the state transition probability in the measurement process for all combinations of $2^n$ bitstrings of quantum states and thus sized $2^n\times 2^n$.
Although the information of correlated errors among qubits due to the leakage of the measurement pulse is also included in addition to the local bit flipping on each qubit, making the complete calibration matrix is not scalable in terms of both the matrix size and the measurement cost.

\par Fortunately, on the near-term devices provided by IBM Quantum Experience, the correlated readout errors are shown to be small enough that one can assume the readout error as either local or correlated among limited spatial extent \cite{mooney2021generation}.
Under this tensor product noise model, the $2^n\times 2^n$ sized calibration matrix $A$ for $n$-qubit measurement process can be seen as a tensor product of fractions of small calibration matrices of local qubit blocks.
This noise model may solve the scalability issue in the complete noise model.
The number of measurements to prepare the local calibration matrices are drastically reduced from $2^n$ to $O(n2^k)$ and the memory to store the calibration matrices is also reduced from $O(4^n)$ to $O(n4^k)$, where $k$ is the size of the biggest complete-calibration block.

\par This tensor product noise model is widely used such as in the standard QREM library of Qiskit Ignis and in other recent works \cite{ctmp2020bravyi, mooney2021generation, nation2021}.
We also develop our proposed QREM methods under this noise model.
Hereinafter, for convenience, we assume the measurement error is completely local,
\begin{equation}
A:=\bigotimes_{i=0}^nA^{(i)},
\end{equation}
where $A^{(i)}$ is the $2\times 2$ calibration matrix of qubit $i$.
With the increase of computational complexity according to the size of the locally complete-mitigation block, the following argument is also applicable to the general tensor product noise models with local calibration matrices for the different sizes of local complete readout channel blocks.

\subsection{Problem Setting}
\label{sec:problem_setting}

\par We are considering the following problem for QREM.
An $n$-qubit measurement result is given as a probability distribution $y \in \mathbb{R}^{|S|}$ where $S\subseteq \{0,1\}^n$ is the subspace of all measured bitstrings with nonzero probability.
Assuming the tensor product noise model, we can also get access to the local calibration matrix $A^
{(i)}$ for each qubit $i$.
Then the task of QREM is to find a probability vector $\tilde x_S\in \mathbb{R}^{|S|}$ that is closest to the probability vector $x$ satisfying $y=Ax$, where the subscript $S$ to the vector (or matrix) $x$ emphasize that the elements of $x$ are in the subspace $S$.
It is also possible to modify the problem to find the extended probability distribution with elements in the subspace $S$ and subspace of bitstrings that are distant from $S$ in Hamming distance $d$.

\par To efficiently solve the defined optimization problem, we will take the following three steps. 
For the first step, we apply the reduced inverse calibration matrix $(A^{-1})_S$ to $y$ to get the ``roughly" mitigated vector $x_S \in \mathbb{R}^{|S|}$ (Step 1).
Since this sparsified inverse matrix would not preserve distribution's sum-to-one condition and nonnegativity, we have to make $x_S$ satisfy these requirements.
Therefore, for the next step, we will find a correction vector $\Delta_S$ that adjusts the sum of elements of $x_S=(A^{-1})_Sy$ into one (Step 2).
Here we propose two different ways to prepare the correction vector.
After this step, while obtaining the corrected vector $\hat x_S = x_S + \Delta_S\in \mathbb{R}^{|S|}$, the negative elements still remains in $\hat x_S$.
To handle this, we cancel the negative values in the corrected vector $\hat x_S$ to finally obtain the proper probability distribution $\tilde x_S\in \mathbb{R}^{|S|}$ (Step 3).

\subsection{Step 1: Matrix Inverse \protect\lowercase{}}
\label{sec:step1}

\par Since the calibration data are given by $n$ $2\times 2$ matrices, $A^{(0)}, \cdots, A^{(n-1)}$, each element of the reduced inverse matrix $(A^{-1})_S$ can be computed in $O(n)$ time respectively by the following way:
\myeq{
\left(A^{-1}\right)_{i j}=\prod_{k=0}^{n-1}\left(A^{(k)}\right)_{i(k), j(k)}^{-1},
}
where $i(k)$ is the $k$-th digit in binary representation of index $i$.
The algorithm for this step is described by the algorithm  \ref{alg:mitigate_one_state} in Appendix \ref{sec:appendix_program}.
This process to prepare the reduced inverse matrix requires $O(n|S|^2)$ time and $O(|S|^2)$ memory.
Once we obtained $(A^{-1})_S$, the roughly mitigated vector $x_S$ is computed by the product of $(A^{-1})_S$ and $y$, i.e. $x_S = (A^{-1})_Sy$.
\par
\begin{figure}[htb]
\includegraphics[width=0.95\linewidth]{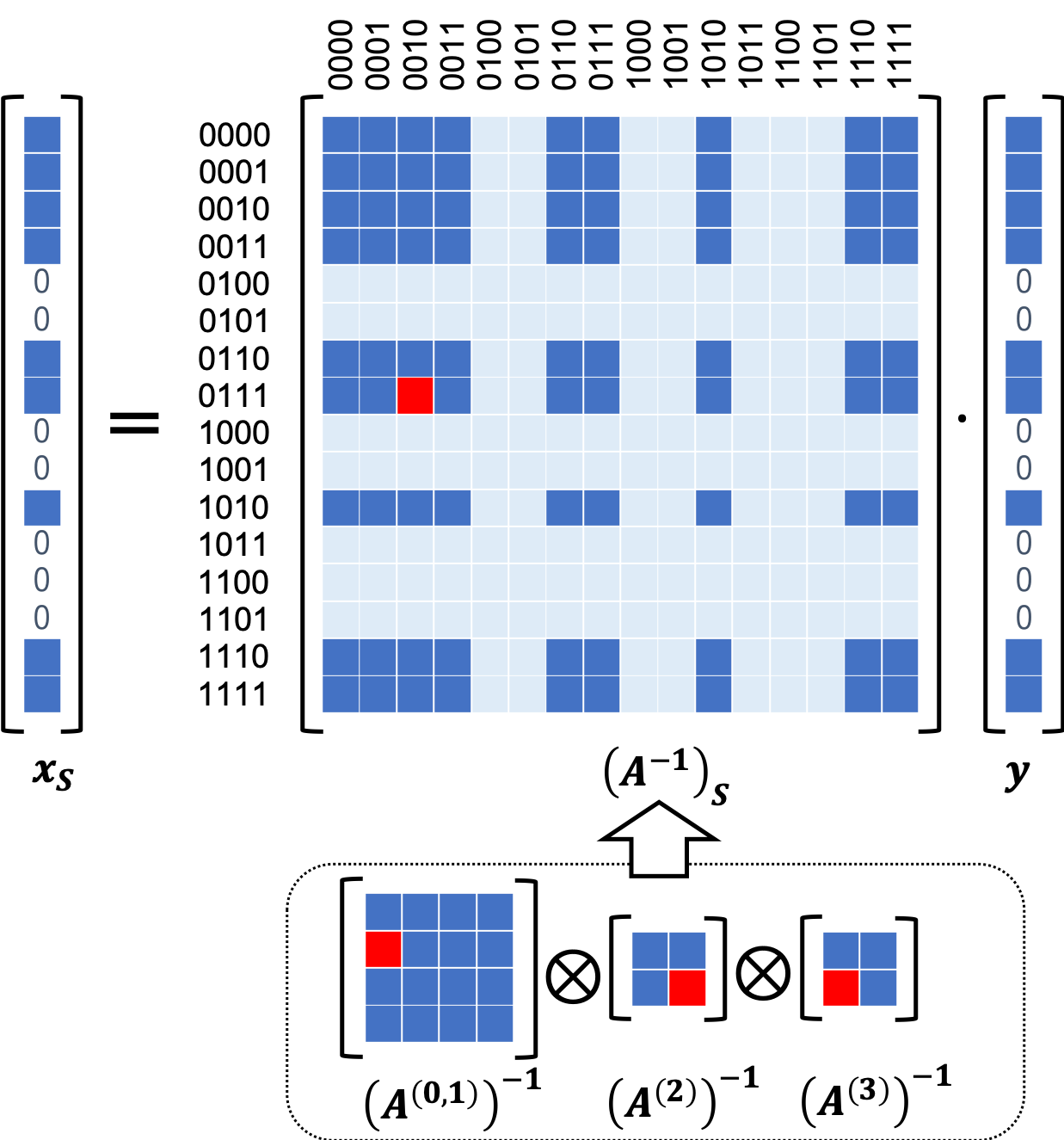}
\caption{\label{fig:step1}
The tensor product noise model and the matrix inversion process of step 1 is shown. The whole calibration matrix $A$ is constructed by the tensor product of local calibration blocks $A^{(0,1)},A^{(2)},A^{(3)}$. For example, the element $A^{-1}_{0111,0010}$ marked in red can be computed by $A^{-1}_{0111,0010} = (A^{(0,1)})^{-1}_{01,00} (A^{(2)})^{-1}_{1,1} (A^{(3)})^{-1}_{1,0}$.
The noisy probability distribution $y$ is assumed to be sparse with nonzero elements in the subspace $S$, which are colored blue.
When computing the roughly mitigated frequency distribution $x_S$, the light-blue colored cells in $A^{-1}$ would not be calculated since their effect are ignored (or drown out) in the subspace $S$ of $y$.
}
\end{figure}

 Note that it is also possible to compute $x_S$ without explicitly preparing $(A^{-1})_S$ as
\myeq{
(x_S)_i = \sum_{j\in S}\prod_{k=0}^{n-1}\left(A^{(k)}\right)_{i(k), j(k)}^{-1}y_j.
}
Then it requires only half the time of making $(A^{-1})_S$ and applying it to $y$, and requires only $O(|S|)$ memory.
In addition, since this step computes the elements of $(A^{-1})_S$ independently, it is compatible with parallel computing frameworks.
The process of step 1 is also depicted in Fig. \ref{fig:step1}.

\par The uncertainty occurring in this step can be evaluated in the same way as mthree \cite{nation2021}.
The mitigation overhead $M$ is determined by the 1-norm of the reduced inverse matrix, $\mathcal M = \|(A^{-1})_S\|_1^2$ \cite{ctmp2020bravyi}.
For an observable $O$, the upper bound on the standard deviation of $O$ becomes
\myeq{
\sigma_O = \sqrt{\mathcal M / s} \label{eq:error_bounds}
}
where $s$ is the number of samples.
Unlike mthree, the mitigation overhead $\mathcal M$ can be rigorously computed because the reduced inverse matrix is explicitly constructed in this step.

\subsection{Step 2: Making the Sum of Vector Elements to One \protect\lowercase{}}
\label{sec:step2}

\par Next is to find a correction vector $\Delta_S$ that makes the element sum of the vector to one.
To compute $\Delta_S$, we first consider the full-sized calibration matrix $A$ and full-sized probability vector $y$ where the empty elements are set to zero.
Let $\hat x$ be $\hat x = x + \Delta$.
Then $\Delta$ is approximated based on the following least square problem:
\begin{eqnarray}
\label{eq:find_delta}
\underset{\Delta}{\text{minimize}}~~ \|A\hat x - y\|^2 = \|A\Delta\|^2,
\end{eqnarray}
\begin{eqnarray}
\text{subject to}~~1^T \hat x &=& 1.
\end{eqnarray}

\par Here we propose two different approaches, delta method and least norm method.
Though we recommend the least norm method in terms of the time complexity, accuracy, and underlying assumptions, we will further see the delta method would perform better than the least norm method in some cases (see Sec. \ref{sec:modified_grover}).

\subsubsection{Approach 1 (delta) \protect\lowercase{}}

In the first approach, we perform the singular value decomposition (SVD) of $A$ and convert the optimization problem above to the form which is analytically solvable.
Let the SVD of $A$ be $A = U\Sigma V^T = \sum_{i=0}^{N-1} \sigma_iu_iv_i^T$ and represent $\Delta$ as $\Delta = \Delta_{j_0}v_{j_0}+ \Delta_{j_1}v_{j_1}+ \cdots \Delta_{j_{k-1}}v_{j_{k-1}}$ using $k$ right singular vectors $\{v_i\}$ of $A$.
Then the problem (\ref{eq:find_delta}) becomes
\begin{eqnarray}
\begin{aligned}
\min_{\Delta \in \mathbb{R}^N} &~ \sum_{i=j_0}^{j_{k-1}} \sigma_i^2 \Delta_i^2 \\
\mbox{subject to} &~ \sum_{i=j_0}^{j_{k-1}} \left(1^T v_i \right)\Delta_i = 1 - 1^T x~. \label{eq:const_optim}
\end{aligned}
\end{eqnarray}
This constrained least square problem can be rigorously solved by Lagrange multiplier.
Each coefficient of $\Delta = \Delta_{j_0}v_{j_0}+ \Delta_{j_1}v_{j_1}+ \cdots \Delta_{j_{k-1}}v_{j_{k-1}}$ can be computed as
\begin{eqnarray}
\Delta_i = \frac{1 - 1^T x} {\displaystyle \sum_{l=j_0}^{j_{k-1}} \frac{\left(1^T v_l\right)^2} {\sigma_l^2}} \frac{1^T v_i} {\sigma_i^2}.
\end{eqnarray}
Since the calibration matrix $A$ is the tensor product of small matrix $A^{(i)}$ for each qubit $i$, the values $\sigma_i, 1^T v_i$ can be computed in $O(n)$ time using the property of $(U_1\Sigma_1 V_1^T)\otimes(U_2\Sigma_2 V_2^T) = (U_1\otimes U_2)(\Sigma_1\otimes \Sigma_2)(V_1^T\otimes V_2^T)$ for the SVD of two matrices $A_1 = U_1\Sigma_1 V_1^T$ and $A_2 = U_2\Sigma_2 V_2^T$.
Restricting the space of the vector into $S$, the correction vector is approximated as $\Delta_S$.
The time complexity to compute the coefficients $\Delta_i$ is $O(n|S|k)$ with arbitrary parameter $k$.
Although this method may not strictly preserve the sum-to-one condition, we will claim that this delta method exploits the benefits of the tensor product noise model.

\label{sec:assumption2}
\par Furthermore, the gap of readout error probabilities $p(1|0)$ getting state 1 expected 0 and $p(0|1)$ for the vice versa, are getting smaller in current devices.
When assuming $p(1|0)\simeq p(0|1)$, the coefficients of $\Delta$ can be approximated more efficiently.
Now the calibration matrix of each qubit $A_i$ becomes closer to a symmetric matrix, which can be eigendecomposed by Hadamard matrices.
Using the property of column sum of the Hadamard matrix, $1^T v_0\gg1^T v_i$ for $i=1,2,\cdots$.
Then $\Delta_i$ can be computed as
\begin{eqnarray}
\label{eq:approximated_delta}
\Delta'_i = \frac{1 - 1^T x} {\frac{\left(1^T v_0\right)^2} {\sigma_0^2}} \frac{1^T v_i} {\sigma_i^2}~.
\end{eqnarray}
In addition, (\ref{eq:approximated_delta}) implies $\Delta'_0 \gg \Delta'_i$ for $i = 1,2,\cdots$.
Therefore, we used $\Delta'_S = \Delta'_0{v_0}_S$ as a correction vector in numerical simulation.
This step takes $O(n|S|)$ time.

\subsubsection{Approach 2 (least norm) \protect\lowercase{}}


\par In the second approach, we find the nearest vector to $x_S$ by solving the following least norm problem.
\myeq{
\begin{gathered}
\min _{\hat{x}_{S}}\left\|\hat{x}_{S}-x_{S}\right\|^{2}, \\
\text { subject to } \mathbf{1}^{\mathrm{T}} \hat{x}_{S}=1.
\end{gathered}
}
By changing the variable with $z := \hat x_S - x_S$, this optimization problem above can be solved by the well-known constrained least norm problem:
\myeq{
\begin{gathered}
\min _{\hat{x}_{S}}\|z\|^{2}, \\
\text { subject to } 1^{\mathrm{T}} z=1-1^{\mathrm{T}} \hat{x}_{S}.
\end{gathered}
}
The analytical solution of this problem is $z=\left(\frac{1-\mathbf{1}^{\mathrm{T}} {x}_{S}}{|S|}\right) \mathbf{1}$.
Therefore the correction process of the second approach becomes
\myeq{
\hat{x}_{S}=x_{S}+\left(\frac{1-\mathbf{1}^{\mathrm{T}} {x}_{S}}{|S|}\right) \mathbf{1}.
}
This simple process requires only $O(|S|)$ time in the computation of $\mathbf{1}^{\mathrm{T}} {x}_{S}$ and addition of the correction term to $x_S$.

\subsection{Step 3: Negative Cancelling \protect\lowercase{}}
\label{sec:step3}
Finally, we are going to find the closest positive vector to $\hat x_S$ which still satisfies the sum--to--one condition.
In this step, the negative canceling algorithm by Smolin, Gambetta, and Smith \cite{Smolin_2012} (the SGS algorithm) is applied.
Given an input vector whose element sum equals 1 but may contain negative values, this algorithm deletes the negative values and the small positive values, and also shifts the positive values to lower ones based on the bounded-minimization approach using the Lagrange multiplier.
The procedure of SGS algorithm is described at  Algorithm \ref{alg:sgs_algorithm}.
Through this algorithm, the finally mitigated probability vector $\tilde x_S = \text{sgs\_algorithm}(\hat x_S)$ is computed in $O(|S|\log |S|)$ time.
Note that the use of SGS algorithm after the main process of matrix inversion process is also mentioned in \cite{mooney2021generation, nation2021}.


\section{\label{sec:demonstrations}Demonstrations\protect\lowercase{}}

\begin{figure*}[htb]
\includegraphics[width=\linewidth]{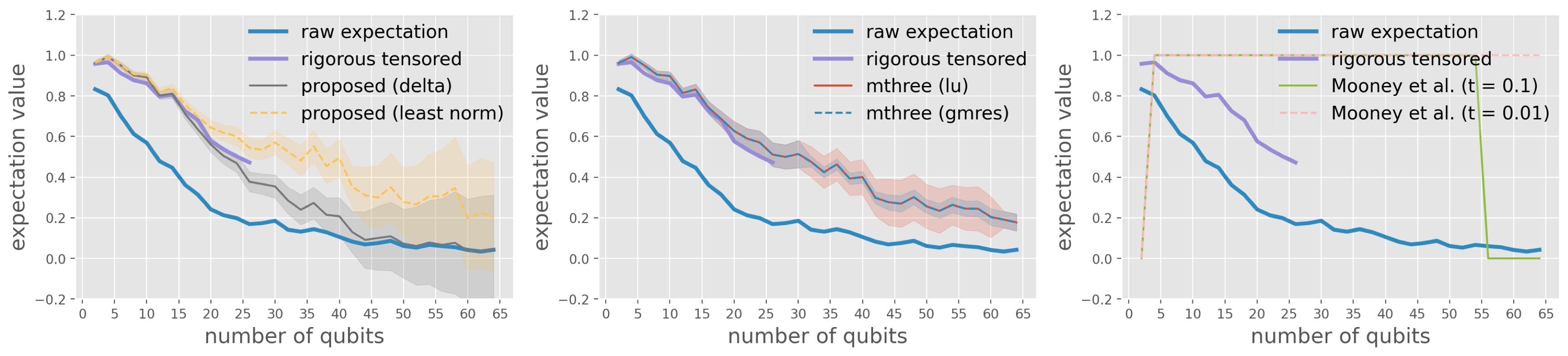}
\caption{\label{fig:ghz_expval}
The plots of expectation values against number of qubits.
All results are normalized to keep the sum of probability vector to one, as in \eqref{eq:expval}.
In the left figure, the solid (gray) line and dashed (yellow) line represent the expectation values by the proposed ``delta" and ``least norm" method introduced in Sec. \ref{sec:propsed_method}.
In the middle figure, the solid (red) line and dotted (blue) line represent the expectation values mitigated by mthree package with ``direct" method using LU decomposition and ``iterative" method using GMRES.
In the right figure, the solid (green) line and dashed (pink) line represent the mitigated expectation values by Mooney \textit{et al.} with threshold $t=0.1$ and $t=0.01$.
As a baseline, all figures contain the expectations with rigorous error mitigation and without error mitigation (raw expectation).
The shaded areas represent error bounds evaluated by the mitigation overhead \eqref{eq:error_bounds}.
The error bounds in the middle figure are computed by the method in mthree.
}
\end{figure*}

\begin{figure*}[htb]
\includegraphics[width=\linewidth]{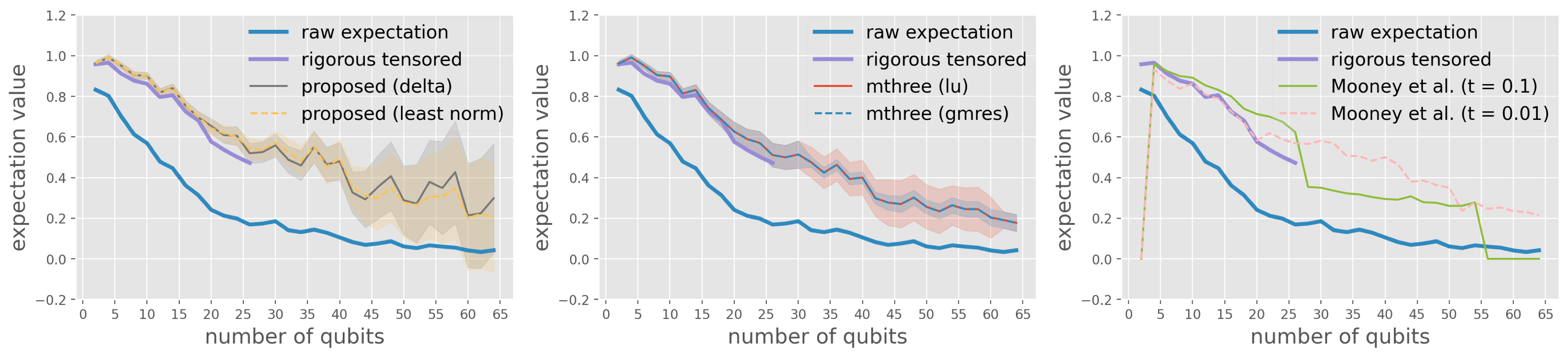}
\caption{\label{fig:ghz_expval_alter}
The plots of expectation values against number of qubits with the normalization method \eqref{eq:expval_alter} instead of \eqref{eq:expval}.
The color notations of plots are the same as those of Fig. \ref{fig:ghz_expval}.
}
\end{figure*}

\par The proposed method is implemented with C++/Eigen, and Cython, and makes use of Qiskit for calibration circuit construction and execution.
The source code of the proposed methods is open to the public named ``libs\_qrem" \cite{libs_qrem}.
For the fair comparison of the performance with existing QREM approaches, both the QREM methods of rigorous inversion of tensor calibration matrix and the method by Mooney \textit{et al.} \cite{mooney2021generation} are also implemented there.
Since the source code of the QREM method by Nation \textit{et al.} \cite{nation2021} is available online, named ``mthree" \cite{mthree}, we use their own implementation for comparison.
We also applied the SGS algorithm \cite{Smolin_2012} to the outputs of mthree to remove the negative elements in the returned probability vectors.
All timing data is taken on 2.5GHz quad-core Intel Core i7 processor (Turbo Boost up to 3.7GHz) with 6MB shared L3 cache with 16GB of 1600MHz DDR3L onboard memory.

\par We conducted two demonstrations on the 65-qubit IBM Quantum Brooklyn system: the expectation values of GHZ states and the fidelity of GHZ states.
The average assignment and CNOT error rates across the qubits on IBM Quantum Brooklyn are 2.89\% and 2.57\% respectively.
The detailed noise information and the mapping of logical qubits to physical qubits are shown in Fig. \ref{fig:ibmq_brooklyn} in Appendix \ref{sec:appendix_device}.
We also conducted the numerical simulation of the modified Grover algorithm on the noisy Qiskit simulator with readout noise to investigate the precision of  estimation error with different QREM methods.

\subsection{Expectation Value of GHZ States on IBM Quantum Brooklyn}

\par First, the mitigated expectation values of GHZ states on IBM Quantum Brooklyn are examined, which is also used as the benchmarking for QREM in \cite{nation2021}.
The expectation value of the observable $O$ is computed by the following way.
Given a raw or mitigated frequency distribution $p$ as a dictionary of bitstrings to their probabilities, the expectation $\braket{O}$ is computed by
\myeq{
\braket{O}=\frac{1}{\operatorname{sum}(p)} \sum_{i} O(i) p_{i}
\label{eq:expval}
}
where $\operatorname{sum}(p)$ is the sum of elements in $p$, $O(i)$ is the value of observable $O$ for state $i$, and $p_i$ is the element in $p$ for the number of shots for state $i$.
Here the expectation value is divided by the sum of elements in $p$ because the mitigated vector might not satisfy the condition of $\operatorname{sum}(p) = 1$.

\par We took the expectation value with the observable $O = \sigma_Z^{(0)}\otimes\cdots\otimes\sigma_Z^{(n-1)}$, measuring all the qubits in computational bases.
The expectations of GHZ state with the even number of qubits are supposed to be 1 under the noiseless environment.


\par Figure \ref{fig:ghz_expval} shows the actual expectation value of GHZ states on IBM Quantum Brooklyn with and without the QREMs.
The mitigation of rigorous inversion of calibration matrix method was performed up to 26-qubit states.
Other efficient QREM methods are performed totally up to 65-qubit measurement results.
The shaded regions give the error bounds by mitigation process which is computed by the mitigation overhead and the number of samples \cite{ctmp2020bravyi}.

\par The left figure in Fig.~\ref{fig:ghz_expval} extracts the plots of two proposed methods (``delta", ``least norm") and the rigorously mitigated expectation values after applying the SGS algorithm.
Expectations by the least norm method are clearly higher than the rigorous inversion of tensor calibration matrix, while the expectations by delta method are closer to the plots of rigorous inversion.
Note that the 1-norm of the reduced inverse matrix $(A^{-1})_S$ used in the computation of error bounds is directly and exactly computed by Eq. (\eqref{eq:error_bounds}).

\par
\begin{figure}[htb]
\includegraphics[width=\linewidth]{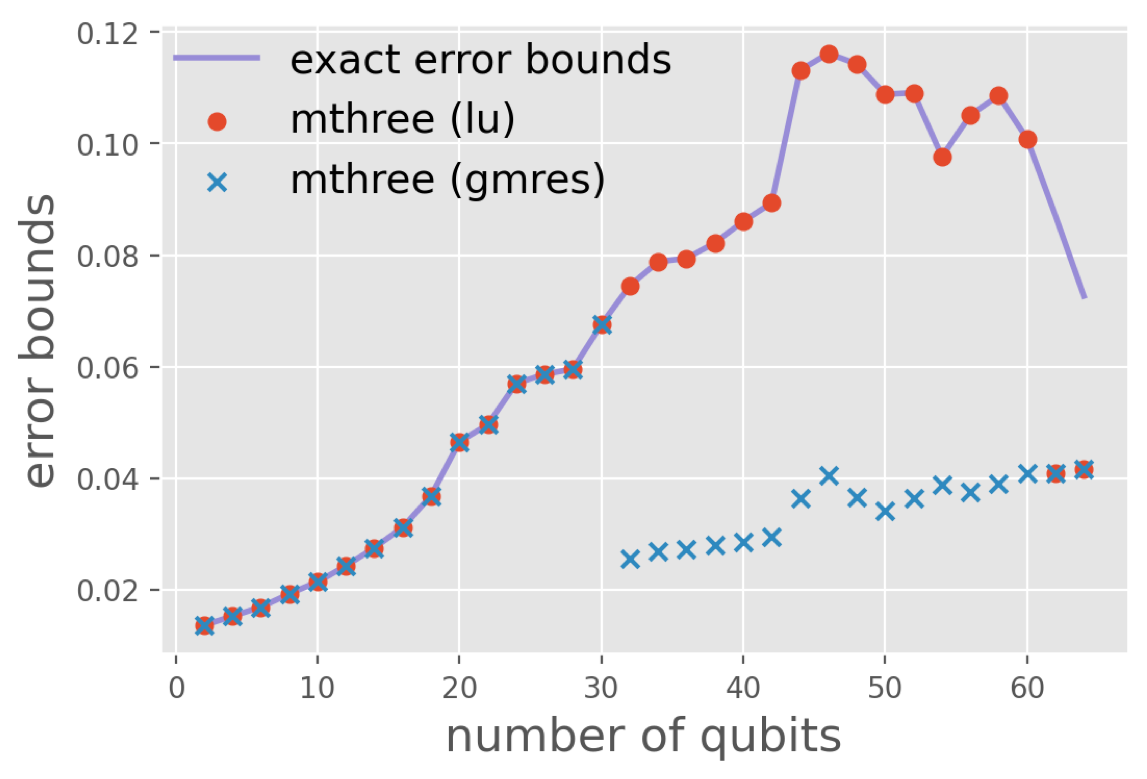}
\caption{\label{fig:error_bounds}
The upper bound on the standard deviation for the expectation of GHZ states with the even number of qubits.
The label ``exact error bounds" denotes the error bounds directly computed from Eq. \eqref{eq:error_bounds}.
The label ``mthree~(lu)" denotes the error bounds computed by Harger's algorithm using LU decomposition in the mthree package.
The label ``mthree~(gmres)" denotes the error bounds computed by Harger's algorithm with GMRES iterative method in the mthree package.
}
\end{figure}

 On the other hand, the central figure in Fig. \ref{fig:ghz_expval} shows the plots of expectation values with QREM by the mthree packages.
Their expectation values are also higher than the rigorously mitigated ones to the same extent as the plots by the proposed ``least norm" method.
Since the inversion of $(A_S)^{-1}$ requires $O(|S|^3)$ times, one can get access only to the approximated error bound through the mthree package for large quantum states.
In addition, the estimated error bound may not always correspond to the exact error bound.
The error bounds through exact computation of $\|(A_S)^{-1}\|_1$ and the error bounds in the mthree package through the iteratively approximated $\|(A_S)^{-1}\|_1$ by Higham's implementation \cite{Higham1988} of Harger's method \cite{Harger1984} are shown in Fig. \ref{fig:error_bounds}.

\par We can also see from the right side figure of Fig. \ref{fig:ghz_expval} that the QREM method of Mooney \textit{et al.} returns the expectation value 1 for almost all the size of GHZ state.
Since the mitigated frequency distributions by this method may not take the element sum to one, the expectation values are normalized by \eqref{eq:expval} as explained above.
On the other hand, we can consider another way to compute expectation value as \eqref{eq:expval_alter}.
\myeq{
\braket{O}=\sum_{i} O(i) p_{i}.
\label{eq:expval_alter}
}
Here we assume the sum of elements of the mitigated frequency distribution is 1, although it may vary through the mitigation process.
This type of expectation values can be seen as the direct counts of the frequency of the bitstrings in the mitigated vector.

\par The expectation values using this calculation method are shown in Fig. \ref{fig:ghz_expval_alter}.
In the right figure of Fig. \ref{fig:ghz_expval_alter}, the mitigated expectation values by the method of Mooney \textit{et al.}, especially with the threshold $0.01$, are close to the rigorously mitigated values.
This can be interpreted as counting the all-zero state and all-one state $\ket{00\ldots 0}, \ket{11\ldots 1}$ in the mitigated vector.
By using Eq. \eqref{eq:expval_alter}, the expectation values of proposed ``delta" method become higher as shown in the left figure of Fig. \ref{fig:ghz_expval_alter}, because the elements of probability distributions by ``delta" method in larger system sizes would no longer sum up to exactly 1.
The gap from 1 becomes even larger for the measurement results of large system while we still use reduced calibration matrix in the small subspace.
Since the ``least norm" method strictly adjust the element sum of the vector to one and mthree uses the quasi-probability in the reduced matrix, the expectation values by these QREM methods take the same values as those in Fig. \ref{fig:ghz_expval}.



\begin{figure}[htb]
\includegraphics[width=\linewidth]{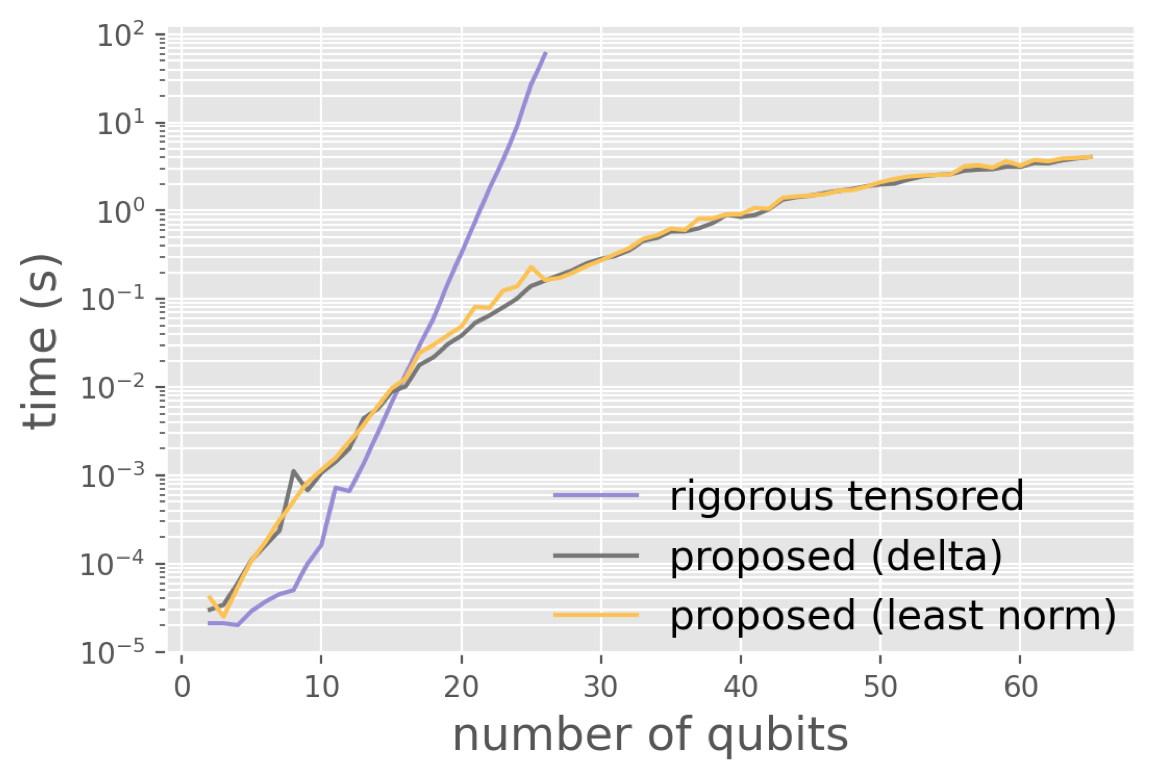}
\caption{\label{fig:times}
The plots of mitigation time by proposed methods and rigorous inversion of tensor calibration matrix.
The mitigation time (y axis) is shown with log scale.
The purple line (with the sharpest slope) denotes the mitigation time by rigorous inversion of tensor calibration matrices.
The black curve denotes the mitigation time by the proposed ``delta" QREM method, and the yellow curve denotes the mitigation time by the proposed ``least norm" QREM method.
}
\end{figure}

\par In addition, the mitigation time of the proposed QREM method is plotted in Fig. \ref{fig:times}.
We can see the rigorous mitigation by the tensor product noise model requires exponential time resources.
Figure \ref{fig:times} implies both of the proposed methods mitigate the 65-qubit GHZ states in 5 s on the 2015 model MacBook Pro, which is practically fast enough for the mitigation of measurement results from large quantum devices.
This high-speed postprocessing also owes to the C++/Cython implementation.

\begin{table*}[htb]
\centering
\caption{\label{tab:ghz_fidelity}
The comparison of measured and mitigated fidelity of GHZ states from 27-qubit to 31-qubit among the rigorous matrix inversion, the proposed ``delta" method, the proposed ``least norm" method, the method of Mooney \textit{et al.} with threshold $0.01$, and the mthree package with ``direct" method using LU decomposition.
}
\begin{tabular}{ccccccc}
\hline\hline
Size & Raw & Rigorous inversion & Proposed (delta) & Proposed (least norm) & Mooney \textit{et al.} ($t=0.01$) & mthree (lu)\\
\hline
27 & $0.344 \pm 0.004$ & $0.541 \pm 0.005$ & $0.420 \pm 0.005$ & $0.543 \pm 0.004$ & $0.580 \pm 0.012$ & $0.520 \pm 0.004$ \\
28 & $0.330 \pm 0.007$ & $0.527 \pm 0.013$ & $0.397 \pm 0.010$ & $0.527 \pm 0.013$ & $0.571 \pm 0.012$ & $0.502 \pm 0.012$ \\
29 & $0.308 \pm 0.004$ & $0.510 \pm 0.008$ & $0.367 \pm 0.006$ & $0.506 \pm 0.008$ & $0.561 \pm 0.007$ & $0.478 \pm 0.007$ \\
30 & $0.299 \pm 0.008$ & $0.491 \pm 0.012$ & $0.353 \pm 0.012$ & $0.494 \pm 0.012$ & $0.560 \pm 0.011$ & $0.464 \pm 0.012$ \\
31 & $0.277 \pm 0.006$ & - & $0.321 \pm 0.009$ & $0.471 \pm 0.011$ & $0.526 \pm 0.013$ & $0.438 \pm 0.010$ \\
\hline\hline
\end{tabular}
\end{table*}

\subsection{Fidelity of GHZ states on IBM Quantum Brooklyn \label{sec:qrem_fidelity}}

\par Next, the fidelity of GHZ states on IBM Quantum Brooklyn was investigated.
We computed the fidelity by multiple quantum coherence (MQC), following the procedure in the demonstrations by Wei et al \cite{Wei2019} and Mooney \textit{et al.} \cite{mooney2021generation}.
The GHZ fidelity $F$ can be calculated as
\myeq{
F:=\frac{P+C}{2}
}
where the population $P = \braket{0\ldots 0|\rho|0\ldots 0} + \braket{1\ldots 1|\rho|1\ldots 1}$ can be directly measured as the GHZ populations and the coherence $C = |\braket{1\ldots 1|\rho|0\ldots 0}| + |\braket{0\ldots 0|\rho|1\ldots 1}|$ can be measured through the MQCs \cite{Wei2019, mqc_baum1985, mqc_garttner2017, mqc_baum1985}.
Here the coherence $C$ is indirectly computed by the following overlap signals $S_\phi = \operatorname{Tr}(\rho_\phi\rho)$, where $\rho_\phi = e^{-i \frac{\phi}{2} \sum_{j} \sigma_{z}^{j}} \rho e^{i \frac{\phi}{2} \sum_{j} \sigma_{z}^{j}}$ is prepared by applying the rotation-Z gates on each qubits.
Using $S_\phi$ with different angle $\phi$, the coherence is calculated as $C = 2\sqrt{I_N}$ with
\myeq{
I_{q}=\mathcal{N}^{-1}\left|\sum_{\phi} e^{i q \phi} S_{\phi}\right|,
}
where $\mathcal N$ is the number of angles $\phi$.


\begin{figure}[htb]
\includegraphics[width=\linewidth]{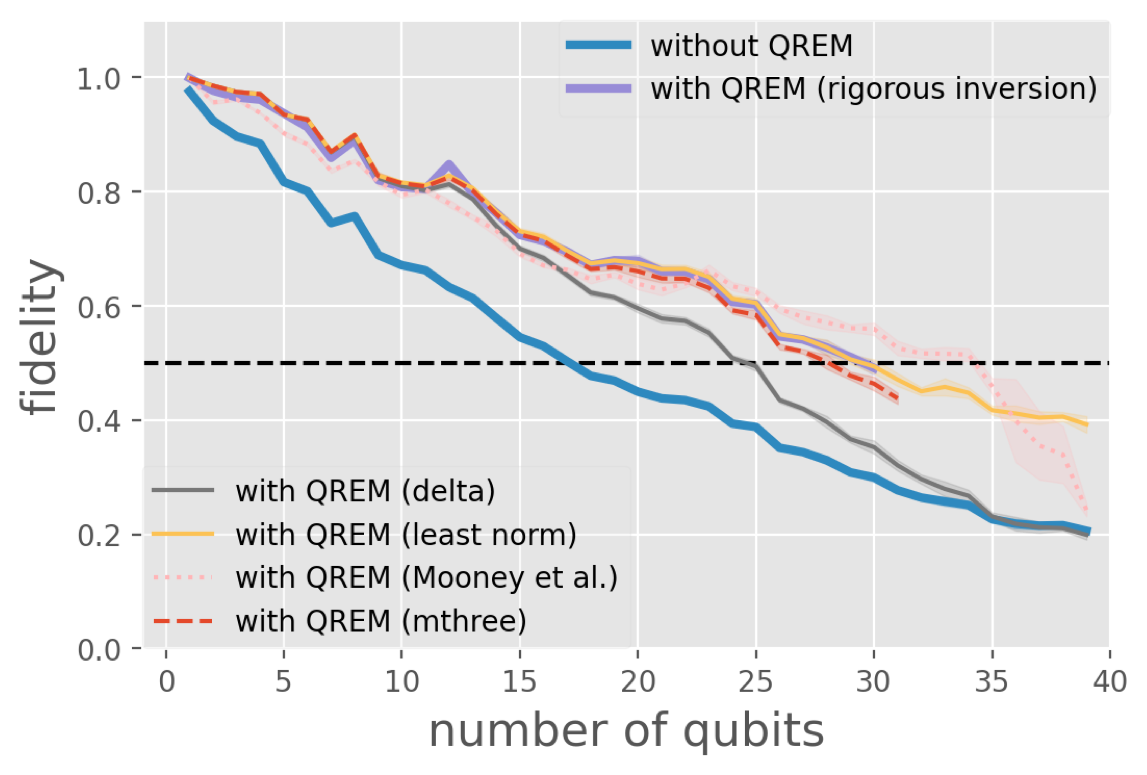}
\caption{\label{fig:ghz_fidelity}
The fidelities of GHZ states on IBM Quantum Brooklyn estimated with different QREM methods.
The blue thick line shows the raw fidelities without QREM.
The purple thick line shows the rigorously mitigated fidelities under the tensor product noise model.
The black line, yellow line, pink dashed line, and red dashed line, respectively, show the fidelities mitigated with the proposed QREM with delta method, the proposed QREM with least norm problem method, the method of Mooney \textit{et al.} with threshold 0.01, and the mthree package with the direct LU decomposition method.
}
\end{figure}

\par The fidelity is averaged over 8 independent runs with 8192 shots as Wei \textit{et al.} \cite{Wei2019} and Mooney \textit{et al.} \cite{mooney2021generation} performed.
Here the fidelity greater than 0.5 is sufficient to confirm the good multipartite entanglement on the real device.

\par The fidelity of GHZ states was examined up to size 39 on the 65-qubit IBM Quantum Brooklyn device.
The different QREM methods are applied to the raw probability distributions.
The QREM with rigorous inversion of tensor calibration matrices is performed up to 30 qubits, while other QREM methods are performed up to 39-qubits.

\par The results are shown in the Fig. \ref{fig:ghz_fidelity}.
The raw results without QREM score the fidelity over 0.5 up to 17-qubits size, while the results with QREM by the proposed method with least square method record the higher fidelities that exceed 0.5 up to qubit size 29 (see Table \ref{tab:ghz_fidelity}).
Note that the fidelity values estimated by the proposed QREM with the least norm problem method closely follow the plots of rigorously mitigated fidelities, while the fidelity plots by other QREM methods have more gaps from the rigorously mitigated fidelities.
The fidelities by the proposed method are also smaller than the fidelities by the method of Mooney \textit{et al.}. From the mitigation results by their method we can observe the 34-qubit GHZ states also scored the fidelity over 0.5 (see Fig. \ref{fig:ghz_fidelity}).

\par Compared to expectation values of the GHZ state, the QREM on the fidelities of the GHZ state seems more effective.
The computation of fidelity uses only the populations of all-zero bitstring and all-one bitstring in the probability distribution, while the computation of expectation value adds up the populations of all the measured bitstrings.
Since the GHZ state outputs only the all-zero and all-one states under the noiseless environment, other bitstrings can be considered as the by-product of various error factors.
However, as the size of the quantum state gets larger, the state preparation error becomes significantly large, which generates more unwanted bistrings in the result probability distribution that would make the computation of expectation value more inaccurate.
Therefore, we can see the readout error mitigation methods for noisy probability distributions are more suitable for recovering the dominant populations in the original probability distribution.

\begin{table*}[htb]
\caption{\label{tab:modified_grover_parameters}
The parameters used in the numerical simulation.
We tried all different condition among the number of qubits, shots for Grover circuits, and the rate readout noise.
The variable $m$ represents the number of Grover iteration in each quantum circuit used in the MLAE algorithm, which is the same among the different conditions.
}
\begin{tabular}{lll}
\hline\hline
Parameter & Parameter name & Examined values \\
\hline 
Number of qubits & $n$ & $\{10, 20\}$ \\
Shots for Grover circuits & $N_{shot}$ & $\{600, 8192\}$ \\
Shots for calibration circuits & - & $\{8192\}$ \\
Number of Grover iteration & $m$ & $[1,2,4,8,16,32,64]$ \\
Target values & $I=\sin^2\theta$ & $b_{max} = \{1/2\}$ \\
Readout noise & $p(0|1)=p(1|0)$ & $\{0.01, 0.03, 0.05\}$ \\
\hline\hline
\end{tabular}
\end{table*}

\begin{figure*}[htb]
\includegraphics[width=\linewidth]{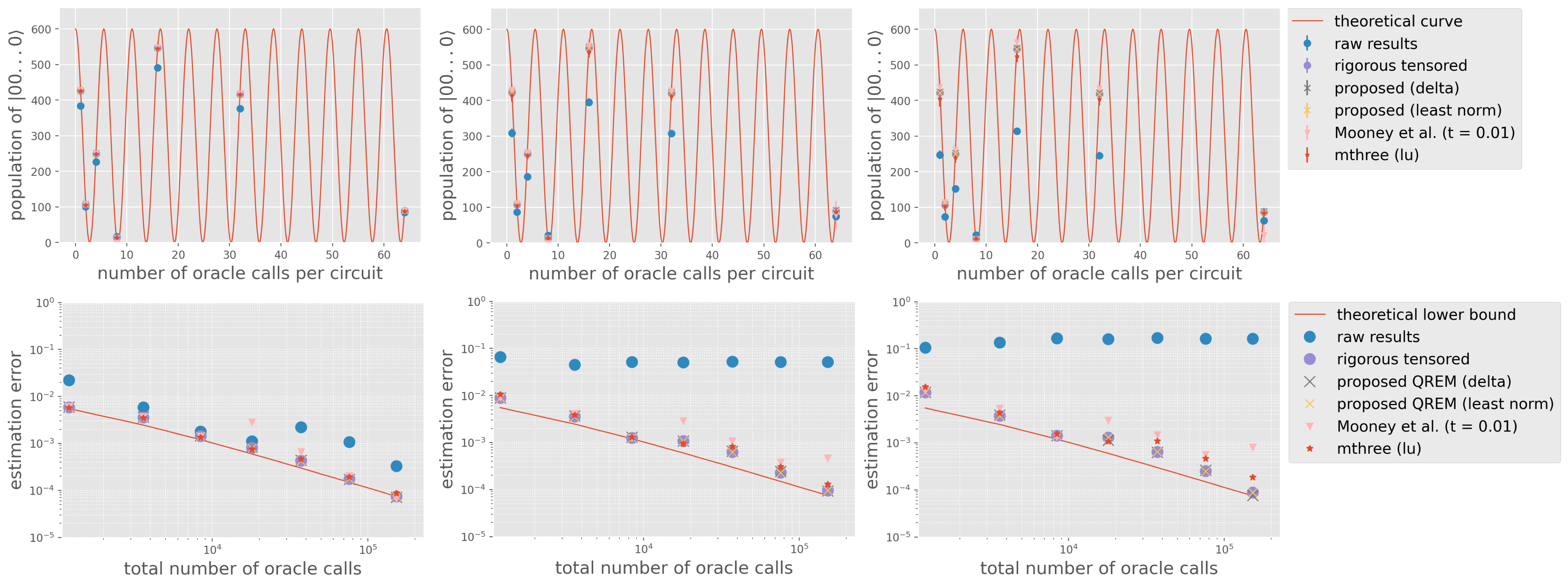}
\caption{\label{fig:10_600}
The number of shots measuring $\ket{0}_{n+1}$ state and the estimation error of Monte Carlo integration on the 10-qubit system where the circuit of Grover iterators are sampled with 600 shots.
The figures in the upper row show the shot count of $\ket{0}_{n+1}$ state from each circuit under different readout noise level.
In the upper row, the red sine waves represent the shot count under the ideal condition without noise.
The other plots represent the measured shot counts with different QREM methods with error bars.
The figures in the lower row show the estimation error of Monte Carlo integration under different readout noise level.
In the lower row, the red lines represent the theoretical lower bound of estimation error (Heisenberg limit).
The other plots represent the estimation errors by different QREM methods.
Figures from left to right are aligned with different noise level of $p(0|1) = p(0|1) = \{0.01, 0.03, 0.05\}$.
}
\end{figure*}

\subsection{Maximum Likelihood Amplitude Estimation with Modified Grover Iterator \label{sec:modified_grover}}

\par Finally, we conduct a Monte Carlo integration by maximum likelihood amplitude estimation (MLAE) algorithm with a modified Grover iterator, which is also called a modified Grover algorithm \cite{uno2020modified}.
This aims to investigate the existence of applications of the proposed method to prospective quantum algorithms.
The whole procedure to estimate the amplitude follows the original MLAE algorithm \cite{Suzuki_2020}, running quantum circuits of a shallower Grover iterator with different iterations.
The MLAE method \cite{Suzuki_2020,Tanaka2021} avoids the phase estimation and many controlled operations in the original amplitude estimation \cite{Brassard_2002} and is expected to be realized earlier than the Shor's algorithm as estimated in~\cite{bouland2020prospects}.
Note that such algorithms without phase estimation were considered folklore~\cite{Aaronson1808}, and have been outlined before, e.g., in~\cite{Grover1998,AW99,PhysRevA.75.012328}.

\par The modified Grover algorithm differs from the MLAE algorithm only in the construction of the Grover iterator, which is represented as $Q=U_{0} A^{\dagger} U_{f} A$, where $U_0$ and $U_f$ are the reflection operators defined as
\begin{eqnarray}
\begin{array}{l}
U_{0}=-\mathbf{I}_{n+1}+2|0\rangle_{n+1}\left\langle\left. 0\right|_{n+1},\right. \\
U_{f}=-\mathbf{I}_{n+1}+2 \mathbf{I}_{n} \otimes|0\rangle\langle 0|.
\end{array}
\end{eqnarray}
The $(n+1)$-qubit initial state $\ket{0}_{n+1}$ after $m$ iterations of operator $Q$ becomes
\begin{eqnarray}
\label{eq:m_iterations}
Q^{m}|0\rangle_{n+1}=\cos \left(2m \theta\right)|0\rangle_{n+1}+\sin \left(2m \theta\right)|\phi\rangle_{n+1}
\end{eqnarray}
where $\ket \phi_{n+1} $ is an unknown state orthogonal to $ \ket 0_{n+1}$.
Then, the probability of getting state $\ket {0}_{n+1}$ with the angle $\theta$ and the number of iteration $m$ is represented as $p_{Q}\left(0 ; \theta, m\right)=\cos ^{2}\left(2 m \theta\right)$.
Hence it is enough to know the probability of getting state $\ket {0}_{n+1}$ in \eqref{eq:m_iterations} to estimate the target value $\theta$.
This type of probability distribution seems to be compatible with applying the proposed QREM algorithm since it is expected to get $\ket {0}_{n+1}$ state with high frequency.

\par According to the MLAE algorithm \cite{Suzuki_2020}, following the Heisenberg limit, the lower bound of the estimation error of $\theta$ decreases at most in the speed of $O(1/\sqrt{N_q})$ for $N_q = N_{shot}\sum_im_i$ queries where $N_{shot}$ is shot count of each quantum circuit.
In contrast, the estimation error converges of in the order of $O(1/N_q)$ for $N_q$ rounds of Grover iterations, which achieves the quadratic speedup.
The performance of the modified Grover algorithm can be checked by the decrease in the rate of estimation errors and how well the estimation errors follow the order of $O(1/N_q)$, which is referred to as the Heisenberg limit.

\par Next, let us briefly review the numerical integration by Grover search following the procedures in \cite{Suzuki_2020}.
Using the notations of \cite{Suzuki_2020}, we focus on the following integration:
\begin{equation}
\begin{aligned}
I &= \frac{1}{b_{\max }} \int_{0}^{b_{\max }} \sin (x)^{2} d x \\
&= \frac 1 {b_{\max}} \left(\frac {b_{\max}} 2 - \frac 1 4 \sin(2b_{\max})\right),
\end{aligned}
\end{equation}
where $b_{\max}$ is an constant parameter. 
The target value $I$ can be discretized as
\begin{equation}
S=\sum_{x=0}^{2^{n}-1} p(x) \sin ^{2}\left(\frac{\left(x+\frac{1}{2}\right) b_{\max}}{2^{n}}\right),
\end{equation}
which can be estimated via the amplitude estimation algorithm and thus the modified Grover algorithm is applicable.

\par We run the numerical simulation of this modified Grover algorithm on the Qiskit simulator.
The simulation was performed with $10$-qubit and $20$-qubit search space, respectively, on the Qiskit simulator \cite{Qiskit}.
The parameter of the modified Grover algorithm following the notation in \cite{uno2020modified} is shown in Table \ref{tab:modified_grover_parameters}.
Since the current QV32 IBM Quantum devices has average readout assignment error from $0.02$ to $0.03$, we tested different readout error rates with $p(0|1) = p(1|0) = \{0.01,0.03,0.05\}$.



\par The results of the numerical simulation are shown in Figs. \ref{fig:10_600}, \ref{fig:10_8192}, \ref{fig:20_600}, and \ref{fig:20_8192}.
Since the estimation error properties of different settings between 10-qubit and 20-qubit systems and between 600 shots and 8192 shots generally share the similar features, we show only the plots of the 10-qubit system with 600 shots as Fig. \ref{fig:10_600}.
The other figures can be found in Appendix \ref{sec:appendix_fig}.
In Fig. \ref{fig:10_600}, the upper rows show the numbers of shots getting $\ket{0}_{n+1}$ and the lower rows show the estimation errors.
All the plots are averaged over ten independent trials.
Plots of each color show
the theoretical values (red curves),
the raw results (blue, ``o"),
the rigorously mitigated results (purple, ``o"),
the mitigated results by proposed QREM with delta method (black, ``x"),
the mitigated results by proposed QREM with least norm method (yellow, ``x"),
the mitigated results by the method of Mooney \textit{et al.} with threshold $t=0.01$ (pink, ``v"),
and the mitigated results by  mthree direct method using the mthree package(red, ``*"). 
Since the plots of Mooney \textit{et al.} with threshold $t=0.1$ are so close theirs with threshold $t=0.01$, they are not shown in the figures.
Likewise, the plots of the mthree iterative method are also omitted because they are so close to the plots of its direct method.

\par Again, we focus only on the results of 10-qubit system with 600 shots in Fig. \ref{fig:10_600}.
In Fig. \ref{fig:10_600}, estimation without QREM fails when the readout errors are set to $0.03$ and $0.05$, while the estimation errors by the rigorous mitigation and by the proposed methods successfully decrease, following the Heisenberg limit.
Compared with mthree and the method of Mooney \textit{et al.}, the proposed methods exhibit better estimation accuracy especially under the higher noise level.
In fact, both the population of $\ket{0}_{n+1}$ and the estimation error by the proposed methods are closer to the rigorously mitigated plots, which is more obvious with 8192 shots (see Figs. \ref{fig:10_8192}, \ref{fig:20_600}, and \ref{fig:20_8192} in Appendix \ref{sec:appendix_fig}).

\par Through these results, it can be said that the modified Grover algorithm is more compatible with the proposed QREM methods to mitigate the readout error.
Also, while the modified Grover algorithm \cite{uno2020modified} is tolerant to the depolarizing noise, these simulation results also support the ability of the modified Grover algorithm to overcome the readout noise.


\section{\label{sec:conclusion}Conclusion\protect\lowercase{}}

\par The proposed QREM methods mitigate the readout error in $O(ns^2)$ time and $O(s)$ memory with $n$ qubits and $s$ shots through the postprocessing on classical computers.
This means the proposed methods scale linearly to the number of finally measured qubits for fixed shot counts, which provides a scalable QREM tool for the current and near-future quantum devices with larger qubits.

\par The demonstrations of GME of GHZ states on IBM Quantum Brooklyn and the numerical simulations of modified Grover algorithm support the advantage of the proposed QREM methods.
The proposed QREM methods mitigate the expectation values of 65-qubit GHZ states on IBM Quantum Brooklyn with exact mitigation overhead while other existing QREM methods can output only the approximated mitigation overhead due to the increase of the computational complexity.
Using the proposed QREM methods, we also witnessed the 29-qubit multipartite entanglement of GHZ state on IBM Quantum Brooklyn with fidelity $0.506 \pm 0.008$.

\par The numerical simulation on the modified Grover algorithm also supports the advantage of the proposed QREM methods.
The estimation errors of target value $\theta$ under different readout noise levels on the 10-qubit system and 20-qubit system are investigated.
In these settings, the proposed methods record the best accuracy among the recently proposed efficient QREM methods.
Therefore, we are likely to find effective applications of the proposed QREM method on such significant quantum algorithms that would be realizable in the near future.
In addition, it can be conversely said that the proposed QREM methods provide a solution to run the modified Grover algorithm under readout noise, revealing the additional advantage of the modified Grover algorithm which is already tolerant to the depolarizing noise.

\begin{acknowledgments}
We thank Prof. Hiroshi Imai at the Graduate School of Information Science and Technology, The University of Tokyo, for the insightful, related discussions and comments.
The results presented in this paper were obtained in part using an IBM Quantum computing system as part of the IBM Quantum Hub at The University of Tokyo.
\end{acknowledgments}

\vspace{0.5\baselineskip}

\textit{Note added.}---
While completing this work, we became aware of a related paper, Ref.~\cite{nation2021}, which was developed independently and concurrently with ours.
The independence and concurrency of our work are demonstrated by our prior conference presentations~\cite{aqis2021,sigqs03}.


\bibliography{main}


\onecolumngrid
\newpage
\appendix

\section{Program Units Used in the Proposed Methods \label{sec:appendix_program}}

The pseudo-code of step 1 in the proposed QREM methods is given by Algorithm \ref{alg:mitigate_one_state}.
This is the most time-consuming step with $O(n|S|^2)$ time and $O(|S|^2)$ memory to the number of qubit $n$ and the dimension of the reduced calibration matrix $|S|$ for the subspace $S\subseteq \{0,1\}^n$.
For $|S|$, we can practically assume $|S| = s$ for shot count $s$.
\begin{figure}[htb]
  \begin{algorithm}[H]
    \caption{Computation of $A^{-1}$}
    \label{alg:mitigate_one_state}
    \begin{algorithmic}
        \Require $m$ local calibration matrices $A^{(0)}, \ldots, A^{(m-1)}$ for arbitrary qubit blocks, noisy probability vector $y$ with subspace $S$
        \Ensure inverse matrix $(A^{-1})_S$ reduced into the subspace $S$
        \State $reduction\_table \leftarrow$ 2d list sized ($|S|$, $m$)
        \For {bitstring $s$ in $S$}
            \For{calibration matrix $A^{(k)}$ in $\{A^{(0)},\cdots, A^{(m-1)}\}$}
                \State $reduction\_table[s, k] \leftarrow s$'s index in the subspace of $A^{(k)}$
            \EndFor
        \EndFor
        \For{target state bitstring $t$ in $S$}
            \For{source state bitstring $s$ in $S$}
                \State product $p \leftarrow 1$
                \For{calibration matrix $A^{(k)}$ in $\{A^{(0)},\cdots, A^{(m-1)}\}$}
                    \State index $i \leftarrow$ $reduction\_table[t, k]$
                    \State index $j \leftarrow$ $reduction\_table[s, k]$
                    \State $p \leftarrow p \cdot A^k_{i,j}$
                \EndFor
                \State $[(A^{-1})_S]_{ij} \leftarrow p$
            \EndFor
        \EndFor
        \State \Return $c$
    \end{algorithmic}
  \end{algorithm}
\end{figure}

Next, Algorithm \ref{alg:sgs_algorithm} shows the method of finding the nearest physically appropriate probability distribution by Smolin, Gambetta, and Smith \cite{Smolin_2012}, which is used in step 3 in the proposed methods.
Our implementation adopts the priority queue to maintain all the elements of frequency distribution $\hat{x}$ in the subspace $S$.
Hence the time complexity of the following pseudo-code is $O(|S|\log|S|)$.

\begin{figure}[htb]
  \begin{algorithm}[H]
    \caption{Negativity Cancellation by Smolin, Gambetta, and Smith \cite{Smolin_2012} (sgs\_algorithm)}
    \label{alg:sgs_algorithm}
    \begin{algorithmic}
        \Require a vector ${\hat x}$ (satisfying $\mathbf{1}^T{\hat x} = 1$) as a dictionary mapping state bitstring to measured count
        \Ensure mitigated probability vector ${\tilde x}$
        \State $queue \leftarrow$ a priority queue of $(value, state)$ sorted with $value$ in ascending order.
        \State accumulator of neative values $neg \leftarrow 0$
        \For{state label $k$ in ${\hat x}$}
            \If {${\hat x}_k < 0$}
                \State $queue.push({\hat x}_k)$
            \EndIf
        \EndFor
        \While {$queue$ is not empty}
            \If {$(neg + queue.top()) / queue.size() < 0$}
                \State $neg \leftarrow neg + queue.pop()$
            \Else
                \State \textbf{break}
            \EndIf
        \EndWhile
        \State mitigated counts ${\tilde x}\leftarrow$ empty vector (dictionary)
        \State division value of negative accumulator $d\leftarrow queue.size()$
        \While {$queue$ is not empty}
            \State ${\tilde x}_k\leftarrow {\hat x}_k + neg / queue.size()$
        \EndWhile
        \State \Return ${\tilde x}$
    \end{algorithmic}
  \end{algorithm}
\end{figure}

\newpage
\section{Estimation Errors of Modified Grover Algorithm \label{sec:appendix_fig}}

The results of numerical simulation of modified Grover algorithm under the 10-qubit system with 8192 shots and 20-qubit system with 600 shots and 8192 shots are shown here.
The shape of these plots are very similar to each other despite the different settings.
We can see the estimations by mthree and the method of Mooney \textit{et al.} are less precise than the proposed methods.

\begin{figure*}[htb]
\includegraphics[width=\linewidth]{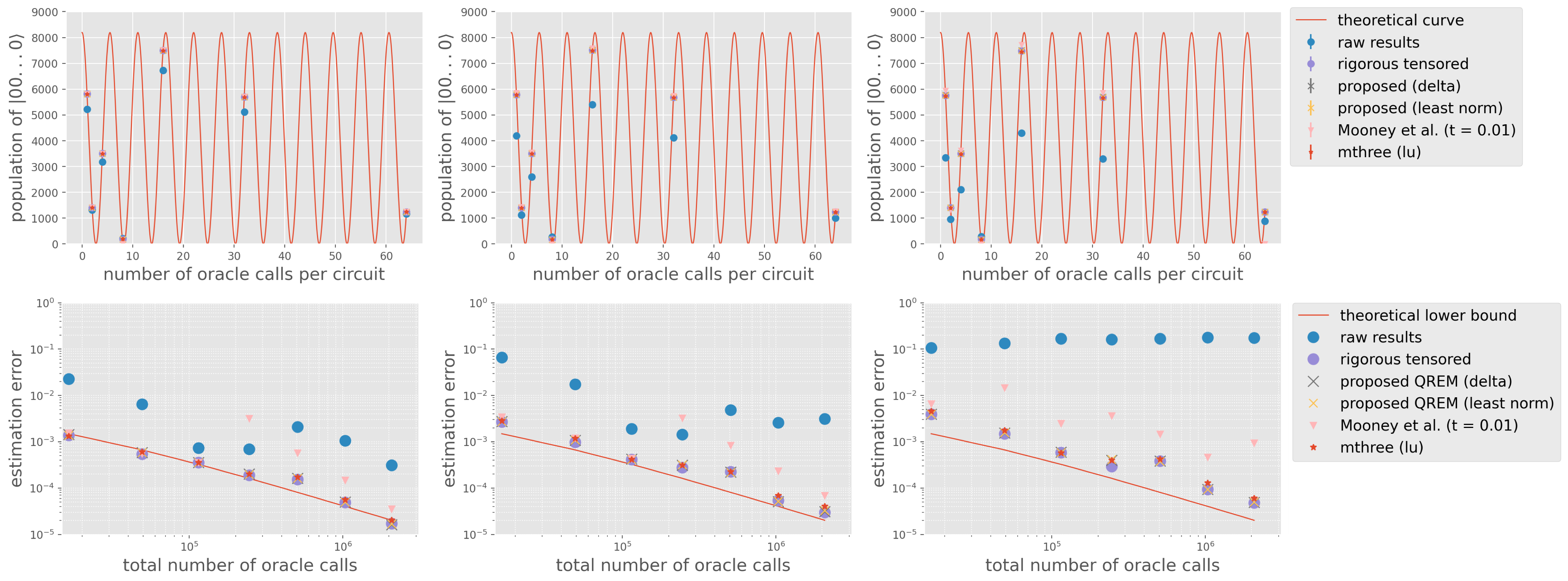}
\caption{\label{fig:10_8192}
The number of shots measuring $\ket{0}_{n+1}$ state and the estimation error of Monte Carlo integration on the 10-qubit system where the circuit of Grover iterators are sampled with 8192 shots.
The figures in the upper row show the shot count of $\ket{0}_{n+1}$ state, and the figures in the lower row show the estimation error of Monte Carlo integration.
Figures from left to right are aligned with different noise level of $p(0|1) = p(0|1) = \{0.01, 0.03, 0.05\}$.
}
\end{figure*}

\begin{figure*}[htb]
\includegraphics[width=\linewidth]{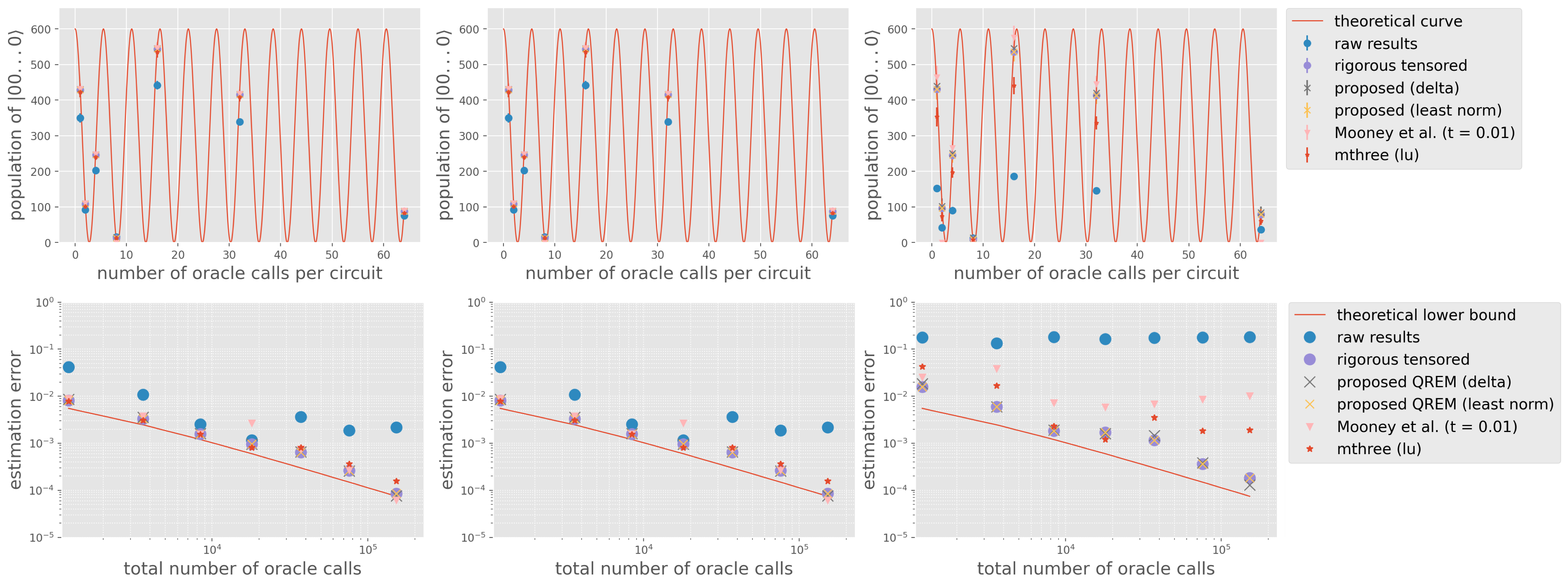}
\caption{\label{fig:20_600}
The number of shots measuring $\ket{0}_{n+1}$ state and the estimation error of Monte Carlo integration on the 20-qubit system where the circuit of Grover iterators are sampled with 600 shots.
The figures in the upper row show the shot count of $\ket{0}_{n+1}$ state, and the figures in the lower row show the estimation error of Monte Carlo integration.
Figures from left to right are aligned with different noise level of $p(0|1) = p(0|1) = \{0.01, 0.03, 0.05\}$.
As the noise level increases, the estimation accuracy by mthree and the method of Mooney \textit{et al.} became clearly lower than the rigorous method and proposed methods.
}
\end{figure*}

\begin{figure*}[htb]
\includegraphics[width=\linewidth]{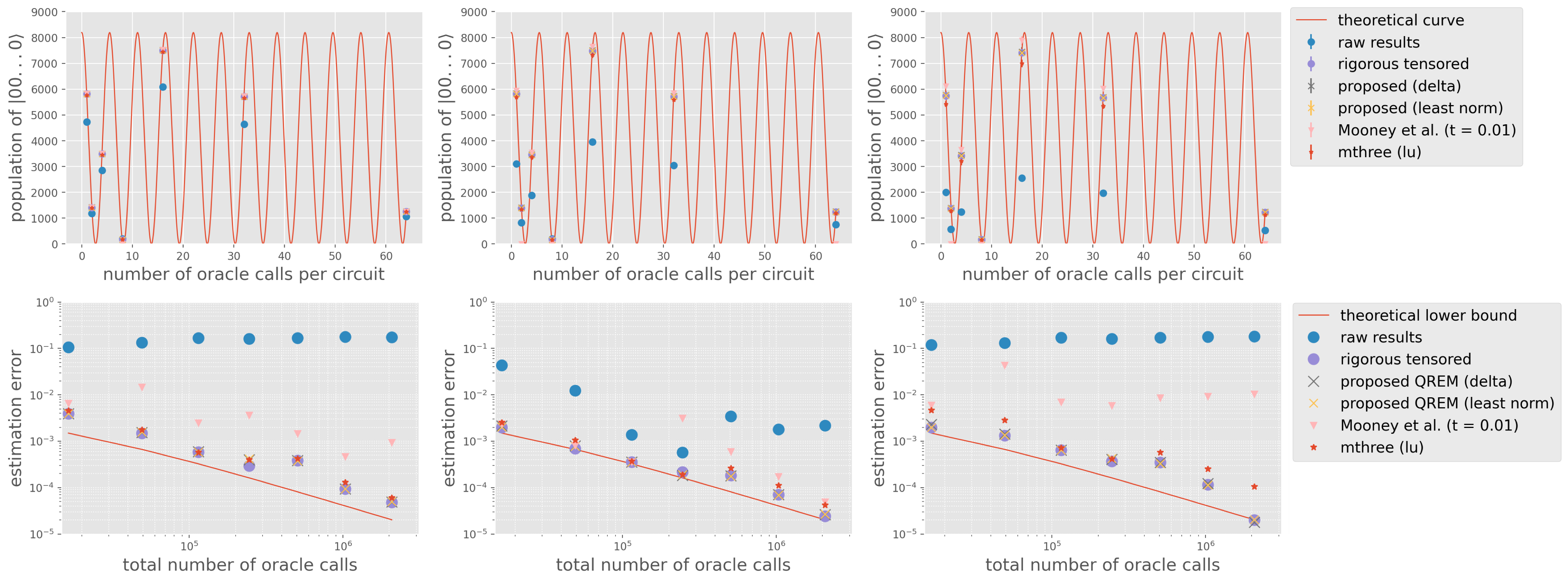}
\caption{\label{fig:20_8192}
The number of shots measuring $\ket{0}_{n+1}$ state and the estimation error of Monte Carlo integration on the 20-qubit system where the circuit of Grover iterators are sampled with 8192 shots.
The figures in the upper row show the shot count of $\ket{0}_{n+1}$ state, and the figures in the lower row show the estimation error of Monte Carlo integration.
Figures from left to right are aligned with different noise level of $p(0|1) = p(0|1) = \{0.01, 0.03, 0.05\}$.
}
\end{figure*}

For all of these settings in Figs. \ref{fig:10_8192}, \ref{fig:20_600}, and \ref{fig:20_8192}, the proposed QREM methods exhibit the best estimation accuracy among other efficient QREM methods \cite{mooney2021generation, nation2021}.
The estimation errors by the proposed methods are the closest to the plots mitigated by the exponential-time rigorous matrix inversion method.

\newpage
\section{Device information of IBM Quantum Brooklyn \label{sec:appendix_device}}

\begin{figure}[htb]
  \begin{minipage}[b]{\linewidth}
    \centering
    \includegraphics[keepaspectratio, width=0.48\linewidth]{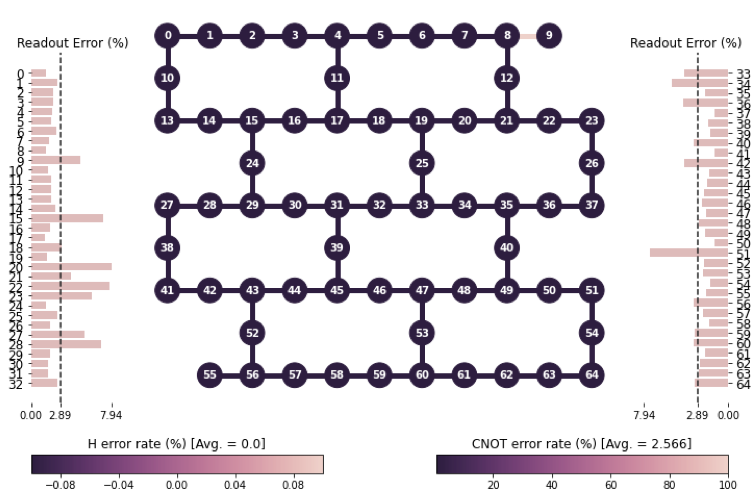}
    \includegraphics[keepaspectratio, width=0.4\linewidth]{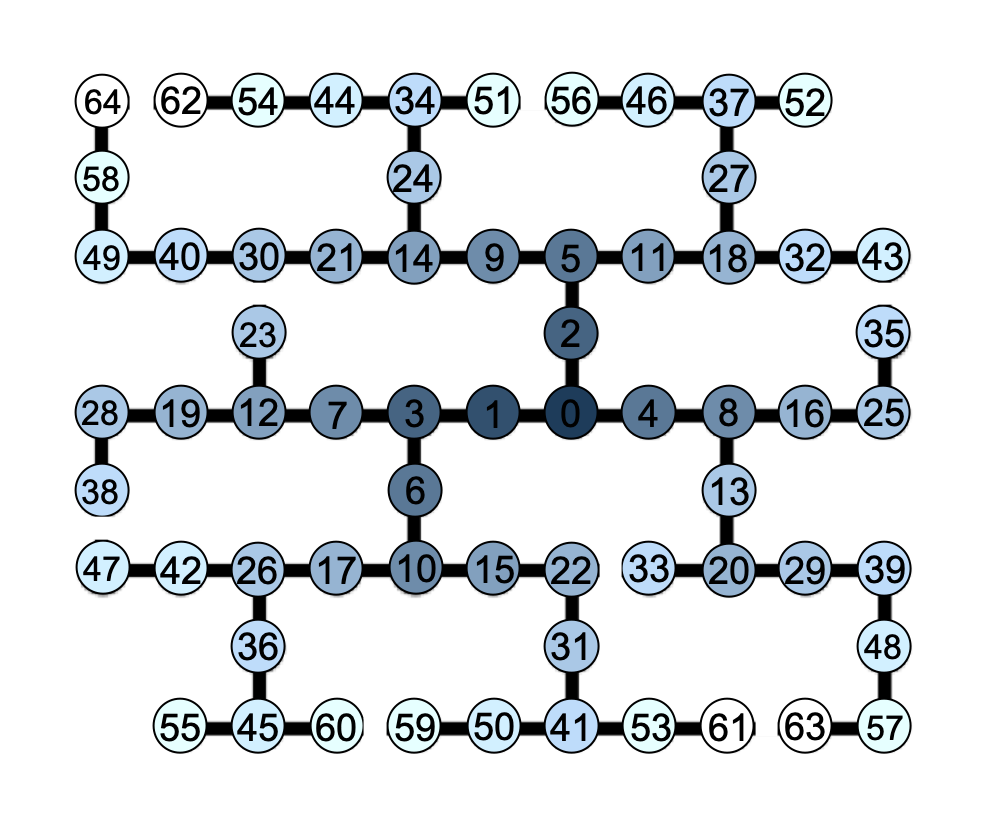}
  \end{minipage}
  \caption{
    \label{fig:ibmq_brooklyn}
    The left figure shows the IBM Quantum Brooklyn error map.
    The numbers on the figure represent the positions of physical qubits.
    More details of noise on each qubit can be found in Table \ref{tab:brooklyn_qubits}.
    The right figure shows the logical qubit layout of GHZ states on IBM Quantum Brooklyn.
    The numbers on the figure represent the positions of logical qubits.
    The size of GHZ states is extended in this order.
    The initial virtual qubits are the physical qubit ``33" to ``32".
    In addition, qubits entangled earlier from the initial qubits are colored with darker blue, and the entangled edges of GHZ states forming the tree structure are colored black.
  }
\end{figure}

The circuits are executed on IBM Quantum Brooklyn by mapping the virtual circuit qubits to physical qubits with [33,
32,
25, 31,
34, 19, 39,
30, 35, 18, 45,
20, 29, 40, 17, 46,
36, 44, 21, 28, 49, 16, 47,
24, 11, 37, 43, 12, 27, 50, 15, 53,
22, 48,  4, 26, 52,  8, 38, 51, 14, 60,
42, 23,  3, 56,  7, 41, 54, 13, 59,
 5,  9, 61,  2, 55,  6, 64, 10, 58,
57, 62,  1, 63, 0]
as shown in Fig. \ref{fig:ibmq_brooklyn}.
The quantum circuit with depth 10 in terms of CNOT gates is enough to prepare a 65-qubit GHZ state using all the qubits in IBM Quantum Brooklyn.
The calibration data of IBM Quantum Brooklyn were retrieved on January 7, 2022.

\begin{table*}[th]
\caption{\label{tab:brooklyn_qubits} Qubit parameters on IBM Quantum Brooklyn. For each qubit, the T1 and T2 relaxation time, frequency and anharmonicity, readout assignment error including the bit-flip probability from 1 to 0 and 0 to 1, and readout length are presented.}
\resizebox{0.9\textwidth}{!}{%
\begin{ruledtabular}
\begin{tabular}{rrrrrrrrr}
Qubit index & T1 (µs) & T2 (µs) & Frequency (GHz) & Anharmonicity (GHz) & Readout Error & $p(0|1)$ & $p(1|0)$ & Readout Length (ns) \\ \hline
0 & 99.687 & 119.304 & 4.853 & -0.334 & 0.015 & 0.024 & 0.006 & 3900.444 \\
1 & 88.170 & 83.400 & 5.004 & -0.331 & 0.027 & 0.037 & 0.016 & 3900.444 \\
2 & 46.516 & 56.531 & 5.146 & -0.328 & 0.023 & 0.034 & 0.011 & 3900.444 \\
3 & 81.080 & 95.507 & 5.050 & -0.329 & 0.022 & 0.030 & 0.014 & 3900.444 \\
4 & 74.301 & 87.578 & 4.938 & -0.330 & 0.021 & 0.030 & 0.012 & 3900.444 \\
5 & 101.850 & 109.505 & 5.010 & -0.331 & 0.020 & 0.028 & 0.012 & 3900.444 \\
6 & 72.008 & 96.854 & 5.133 & -0.328 & 0.025 & 0.041 & 0.009 & 3900.444 \\
7 & 84.919 & 113.334 & 4.975 & -0.331 & 0.018 & 0.024 & 0.012 & 3900.444 \\
8 & 76.755 & 82.536 & 5.164 & -0.328 & 0.015 & 0.022 & 0.007 & 3900.444 \\
9 & 74.274 & 20.933 & 5.305 & -0.325 & 0.049 & 0.076 & 0.023 & 3900.444 \\
10 & 84.241 & 113.813 & 5.065 & -0.329 & 0.017 & 0.026 & 0.008 & 3900.444 \\
11 & 65.223 & 82.845 & 5.085 & -0.327 & 0.020 & 0.033 & 0.008 & 3900.444 \\
12 & 96.798 & 111.944 & 5.017 & -0.329 & 0.020 & 0.025 & 0.014 & 3900.444 \\
13 & 66.575 & 12.480 & 5.256 & -0.326 & 0.020 & 0.026 & 0.014 & 3900.444 \\
14 & 47.583 & 56.249 & 5.203 & -0.327 & 0.024 & 0.037 & 0.011 & 3900.444 \\
15 & 56.742 & 65.245 & 5.073 & -0.329 & 0.072 & 0.130 & 0.014 & 3900.444 \\
16 & 75.892 & 116.687 & 5.296 & -0.326 & 0.019 & 0.028 & 0.010 & 3900.444 \\
17 & 67.538 & 72.163 & 5.221 & -0.326 & 0.014 & 0.022 & 0.006 & 3900.444 \\
18 & 88.163 & 82.514 & 5.276 & -0.326 & 0.030 & 0.045 & 0.015 & 3900.444 \\
19 & 88.271 & 89.443 & 5.077 & -0.329 & 0.015 & 0.022 & 0.009 & 3900.444 \\
20 & 34.346 & 27.094 & 4.940 & -0.324 & 0.079 & 0.118 & 0.040 & 3900.444 \\
21 & 79.739 & 83.384 & 5.117 & -0.330 & 0.040 & 0.061 & 0.019 & 3900.444 \\
22 & 46.978 & 52.597 & 5.034 & -0.330 & 0.078 & 0.142 & 0.015 & 3900.444 \\
23 & 83.248 & 74.656 & 4.869 & -0.333 & 0.060 & 0.066 & 0.055 & 3900.444 \\
24 & 84.121 & 110.114 & 5.299 & -0.326 & 0.015 & 0.022 & 0.008 & 3900.444 \\
25 & 91.718 & 122.664 & 5.159 & -0.327 & 0.026 & 0.041 & 0.012 & 3900.444 \\
26 & 82.130 & 54.040 & 5.093 & -0.328 & 0.018 & 0.030 & 0.007 & 3900.444 \\
27 & 15.231 & 24.290 & 5.315 & -0.332 & 0.053 & 0.084 & 0.023 & 3900.444 \\
28 & 65.363 & 14.242 & 5.270 & -0.325 & 0.070 & 0.080 & 0.059 & 3900.444 \\
29 & 56.699 & 68.153 & 5.210 & -0.327 & 0.020 & 0.028 & 0.011 & 3900.444 \\
30 & 88.793 & 28.484 & 5.083 & -0.329 & 0.017 & 0.026 & 0.009 & 3900.444 \\
31 & 48.444 & 46.492 & 5.195 & -0.327 & 0.017 & 0.027 & 0.008 & 3900.444 \\
32 & 78.048 & 107.926 & 5.401 & -0.322 & 0.026 & 0.041 & 0.012 & 3900.444 \\
33 & 73.209 & 103.080 & 5.258 & -0.326 & 0.042 & 0.058 & 0.025 & 3900.444 \\
34 & 93.767 & 137.213 & 5.054 & -0.329 & 0.053 & 0.063 & 0.044 & 3900.444 \\
35 & 101.154 & 113.431 & 4.814 & -0.334 & 0.022 & 0.035 & 0.010 & 3900.444 \\
36 & 86.391 & 118.915 & 5.066 & -0.330 & 0.042 & 0.058 & 0.026 & 3900.444 \\
37 & 73.752 & 110.201 & 5.187 & -0.327 & 0.013 & 0.020 & 0.007 & 3900.444 \\
38 & 88.681 & 93.934 & 5.096 & -0.329 & 0.019 & 0.026 & 0.012 & 3900.444 \\
39 & 59.262 & 84.387 & 5.307 & -0.326 & 0.016 & 0.021 & 0.012 & 3900.444 \\
40 & 108.386 & 110.879 & 4.930 & -0.331 & 0.033 & 0.054 & 0.011 & 3900.444 \\
41 & 67.587 & 90.364 & 5.214 & -0.327 & 0.013 & 0.021 & 0.005 & 3900.444 \\
42 & 69.936 & 87.408 & 5.286 & -0.326 & 0.042 & 0.060 & 0.024 & 3900.444 \\
43 & 49.758 & 71.360 & 5.113 & -0.328 & 0.018 & 0.026 & 0.010 & 3900.444 \\
44 & 85.050 & 41.332 & 5.291 & -0.326 & 0.020 & 0.026 & 0.015 & 3900.444 \\
45 & 84.377 & 52.278 & 5.070 & -0.329 & 0.022 & 0.035 & 0.010 & 3900.444 \\
46 & 73.369 & 91.760 & 5.133 & -0.328 & 0.024 & 0.037 & 0.011 & 3900.444 \\
47 & 73.567 & 89.407 & 5.263 & -0.325 & 0.021 & 0.027 & 0.014 & 3900.444 \\
48 & 75.969 & 86.739 & 5.315 & -0.326 & 0.027 & 0.036 & 0.019 & 3900.444 \\
49 & 70.735 & 93.343 & 5.151 & -0.328 & 0.022 & 0.035 & 0.009 & 3900.444 \\
50 & 61.920 & 65.636 & 5.059 & -0.329 & 0.013 & 0.018 & 0.008 & 3900.444 \\
51 & 51.111 & 54.139 & 5.246 & -0.326 & 0.074 & 0.135 & 0.014 & 3900.444 \\
52 & 77.455 & 128.381 & 5.031 & -0.330 & 0.023 & 0.038 & 0.008 & 3900.444 \\
53 & 84.407 & 101.903 & 5.053 & -0.329 & 0.024 & 0.034 & 0.013 & 3900.444 \\
54 & 80.380 & 89.506 & 5.145 & -0.328 & 0.016 & 0.027 & 0.006 & 3900.444 \\
55 & 77.412 & 118.046 & 5.201 & -0.326 & 0.021 & 0.031 & 0.011 & 3900.444 \\
56 & 49.125 & 44.026 & 5.243 & -0.327 & 0.033 & 0.050 & 0.015 & 3900.444 \\
57 & 89.308 & 95.953 & 5.014 & -0.330 & 0.024 & 0.034 & 0.014 & 3900.444 \\
58 & 82.202 & 94.771 & 5.165 & -0.328 & 0.018 & 0.029 & 0.008 & 3900.444 \\
59 & 64.740 & 60.478 & 5.003 & -0.329 & 0.031 & 0.050 & 0.013 & 3900.444 \\
60 & 64.236 & 69.999 & 5.118 & -0.328 & 0.033 & 0.050 & 0.015 & 3900.444 \\
61 & 86.322 & 92.857 & 5.260 & -0.326 & 0.022 & 0.029 & 0.015 & 3900.444 \\
62 & 40.405 & 72.328 & 4.998 & -0.329 & 0.027 & 0.043 & 0.010 & 3900.444 \\
63 & 72.136 & 67.759 & 5.086 & -0.329 & 0.027 & 0.040 & 0.014 & 3900.444 \\
64 & 35.479 & 50.454 & 5.365 & -0.325 & 0.031 & 0.040 & 0.023 & 3900.444 \\
\end{tabular}
\end{ruledtabular}}
\end{table*}

\end{document}